\documentclass[sigconf]{acmart}
\AtBeginDocument{%
  }

\usepackage{multirow}
\usepackage{subcaption}
\usepackage{rotating}
\usepackage{microtype}
\usepackage{colortbl}

\copyrightyear{2026}
\acmYear{2026}
\setcopyright{cc}
\setcctype{by}
\acmConference[RecSys '26]{20th ACM Conference on Recommender Systems}{September 27-October 02, 2026}{Minneapolis, MN, USA}
\acmBooktitle{20th ACM Conference on Recommender Systems (RecSys '26), September 27-October 02, 2026, Minneapolis, MN, USA}
\acmDOI{10.1145/3773078.3831830}
\acmISBN{979-8-4007-2284-4/2026/09}

\begin{document}

\title{A Redundancy Reduction Approach for Controllable Sequential Recommendations}

\author{Veronika Ivanova}
\orcid{0009-0002-7080-4648}
\affiliation{%
  \institution{Yandex}
  \city{Moscow}
  \country{Russia}
}
\affiliation{%
  \institution{Applied AI Institute}
  \city{Moscow}
  \country{Russia}
}
\email{veronika.ivanova88@yandex.ru}

\author{Marina Munkhoeva}
\orcid{0000-0002-5638-3712}
\affiliation{%
  \institution{AXXX}
  \city{Moscow}
  \country{Russia}
}
\affiliation{%
  \institution{Lomonosov MSU}
  \city{Moscow}
  \country{Russia}
}

\author{Ivan Razvorotnev}
\orcid{0009-0000-4820-8482}
\affiliation{%
  \institution{Applied AI Institute}
  \city{Moscow}
  \country{Russia}
}
\affiliation{%
  \institution{HSE University}
  \city{Moscow}
  \country{Russia}
}

\author{Evgeny Frolov}
\orcid{0000-0003-3679-5311}
\affiliation{%
  \institution{AXXX}
  \city{Moscow}
  \country{Russia}
}
\affiliation{%
  \institution{HSE University}
  \city{Moscow}
  \country{Russia}
}

\renewcommand{\shortauthors}{Ivanova et al.}

\begin{abstract}
Sequential recommendation must operate under long-tailed item distributions and popularity-driven concentration, often forcing practitioners to trade short-list accuracy against long-tail exposure.
In this work, we study \emph{feature decorrelation} as a mechanism for shaping representation geometry in dot‑product sequential recommenders, and analyze how this, in turn, affects popularity‑driven concentration.
We propose a decorrelation-regularized training framework that augments next-item prediction with an auxiliary redundancy-reduction term, and instantiate it with \textbf{BT-SR}, which uses the Barlow Twins objective.
To form label-consistent positive pairs without synthetic corruptions, we pair user histories that share the same next-item target.
Beyond accuracy, we provide a geometric analysis suggesting how decorrelation suppresses shared low-rank directions in the user representation space that may give popular items a global scoring advantage, and we introduce a bucket-based \emph{alignment concentration} metric to quantify this effect.
Experiments on five public benchmarks show that BT-SR consistently improves next-item ranking quality, while the decorrelation strength acts as a simple control knob that reallocates accuracy across head and tail items, enabling accuracy--exposure trade‑offs. Our analysis also reveals that the impact on head‑vs‑tail exposure differs across datasets, reflecting interactions between decorrelation and data temporal structure.
\end{abstract}




\maketitle

\section{Introduction}
\label{sec:intro}
\sloppy
Sequential recommendation models infer a user's next interaction from their recent history and serve as a core component of modern recommender systems.
Transformer-based architectures~\cite{DBLP:journals/corr/VaswaniSPUJGKP17} have become the standard backbone due to their ability to represent complex temporal patterns in user behavior, yet their effectiveness is often limited by extreme interaction sparsity with skewed item popularity distributions.
In practice, deployed systems face intertwined challenges: (i) head-aligned geometric structure, where popular items dominate exposure and reduce catalog coverage; (ii) conflicting objectives such as precision versus diversity; and (iii) the tension between shared short-term intent and user-specific long-term preferences.

A popular approach to combating sparsity is to add self-supervised learning objectives to sequential recommenders, most commonly via contrastive learning~\cite{DBLP:journals/corr/abs-2010-14395, DBLP:journals/corr/abs-2110-05730, du2022contrastivelearningbidirectionaltransformers, liu2021contrastiveselfsupervisedsequentialrecommendation}. Contrastive methods improve accuracy by pulling together two views of a sequence while pushing representations of different sequences apart, but this paradigm introduces additional design and engineering burdens. 
In particular, negative sampling and batch construction interact in complex ways with long-tailed data and can unintentionally reinforce head-item dominance~\cite{cai2024popularityawarealignmentcontrastmitigating, prakash2024evaluating}. Moreover, the effectiveness of contrastive learning often depends on carefully tuned augmentation pipelines and large batches~\cite{DBLP:journals/corr/abs-2010-04592, kulatilleke2022efficientblockcontrastivelearning}, which complicates training and limits controllability over the resulting trade-offs.

One way to enforce the desired trade-offs is to explicitly impose fairness or exposure constraints, as in popularity‑debiasing methods. Our goal is different: we aim at understanding an inherent structure of learned representations and reshaping the geometry of user and item embeddings, which in turn may modulate head--tail exposure. Because our focus is on characterizing this emergent geometric effect rather than proposing a new debiasing mechanism, we do not benchmark against dedicated popularity-debiasing methods; such methods address a different problem setting and are not directly comparable to our representation-geometry analysis. \emph{We treat popularity‑driven concentration as an emergent property of representation geometry, not as a target for direct correction}. We specifically focus on shaping representation geometry via \emph{feature decorrelation}.
Non-contrastive redundancy-reduction objectives (e.g., Barlow Twins~\cite{zbontar2021barlow}, VICReg~\cite{bardes2022vicreg}, whitening-based losses~\cite{huang2018decorrelated, ermolov2021whitening, hua2021feature}) are attractive because they do not require explicit negatives and directly constrain global structure of the embedding space.
However, for sequential recommendation it remains unclear what exactly such objectives do to user and item embeddings, and how we can benefit from these geometric changes.

We study decorrelation as a mechanism for navigating the accuracy--coverage trade-off in dot-product sequential recommenders. Our approach is a \emph{decorrelation-regularized} training framework that augments standard next-item prediction loss with an auxiliary redundancy-reduction term, $L_{\text{pred}} + \alpha L_{\text{decorr}}$. We instantiate this framework with BT-SR, where $L_{\text{decorr}}$ is the Barlow Twins loss. To avoid synthetic sequence corruptions and keep the auxiliary objective label-consistent, we construct positive pairs by sampling distinct user histories that share the same next-item target. This augmentation strategy naturally aligns with the recommendation objective: sequences that lead to the same item should have similar representations, while avoiding heuristics such as cropping or masking that may distort user intent. This yields a simple multi-task training scheme with a single interpretable knob $\alpha$ that continuously influences how decorrelation shapes the learned space.

Beyond empirical gains, we provide a geometric intuition on how decorrelation reshapes the accuracy--coverage trade-off.
We suggest that the tendency for frequent items to dominate rankings can be viewed as alignment of item embeddings with shared low-rank directions in the user representation space. 
Decorrelation suppresses these shared directions by promoting approximately isotropic user representations, thereby reducing the ability of head items to dominate rankings via global alignment.
To quantify this effect, we introduce a bucket-based \emph{Alignment Concentration} metric based on item energy in the top principal subspace of user embeddings, and relate it to head-tail exposure patterns. Our contributions are:
\begin{itemize}
\item a new perspective on why redundancy reduction matters in sequential recommendation: it changes the geometry of user and item representations in a way that influences the head-tail distribution of recommended items;
\item a simple and reproducible instantiation BT-SR that integrates decorrelation into standard next-item training with an interpretable strength parameter;
\item an alignment concentration metric that tracks how strongly the shared structure is concentrated in items with different popularity levels, enabling fine-grained model-by-model comparison across datasets.
\end{itemize}
Our empirical study further reveals that the impact of decorrelation on head--tail exposure is dataset‑dependent: it may either improve or reduce catalog coverage due to intrinsic dataset structure.

\section{Related Work}

\paragraph{Sequential Recommendation with Transformers.}
Transformer-based architectures, particularly SASRec~\cite{DBLP:journals/corr/abs-1808-09781} and BERT4Rec~\cite{sun2019bert4recsequentialrecommendationbidirectional}, have become the dominant paradigm for sequential recommendation due to their ability to model long-range dependencies and encode rich user-history representations. These works established Transformers as a strong default backbone and motivated a broad range of refinements that target both representation quality and training stability.
Subsequent research has explored architectural and optimization choices that affect sequential recommenders, including alternative positional encodings, attention variants, and regularization strategies~\cite{petrov2024transformers}. Parallel work has also examined how the training objective, such as autoregressive or masked modeling, shapes the aspects of user intent captured by the model and how these choices interact with data sparsity and long-tailed item distributions. Our work builds on this Transformer-based foundation but focuses on an orthogonal question: how explicitly shaping representation geometry through feature decorrelation influences recommendation behavior.

\paragraph{Next-Item Objectives and Training Efficiency.}
Another key design axis is the next-item objective. Many Transformer-based recommenders use pairwise or pointwise losses such as BCE with negative sampling~\cite{DBLP:journals/corr/abs-1808-09781, petrov2023gsasrec}, which is efficient but depends on the sampling strategy. Softmax CE can improve ranking quality~\cite{dross2023}, but full normalization over large catalogs is often prohibitively expensive, motivating sampled or approximate alternatives. 
Recent scalable objectives aim to retain the benefits of softmax-style learning while reducing computational overhead. For instance, SCE~\cite{Mezentsev_2024} replaces full-catalog normalization with dynamically sampled candidate sets, enabling efficient training for large-scale recommendation. 

Importantly, objective choice can also influence representation geometry and gradient concentration on frequent items under different normalization schemes, which can affect popularity-driven concentration in practice. We therefore evaluate the decorrelation regularizer across BCE, full softmax CE, and SCE to disentangle representation effects from the choice of prediction loss and to assess the robustness of decorrelation-induced behavior across training regimes.
\paragraph{Self-Supervised Learning in Sequential Recommendation.}
While backbone and objective define the core sequential recommender, recent work increasingly augments these models with auxiliary self-supervised signals to improve representation learning under sparsity and long-tailed data. Many methods follow contrastive learning (CL): they create two augmented views of a sequence and pull them together while pushing representations of different sequences apart. CL4SRec~\cite{xie2021contrastivelearningsequentialrecommendation} systematized common augmentation operators (cropping, reordering, masking, substitution), while DuoRec~\cite{DBLP:journals/corr/abs-2110-05730} proposed target-aware pairing to better preserve user intent, a technique that became common in many contrastive learning schemes \cite{wang2024relative}. EC4SRec~\cite{wang2022explanationguidedcontrastivelearning} is an explanation-guided contrastive framework that leverages training gradients to identify positive and negative items. Extensions such as CBiT~\cite{du2022contrastivelearningbidirectionaltransformers} and SCL~\cite{shi2024self} further refine the training signal through bidirectionality and regularization terms aimed at stabilizing optimization.

Despite strong empirical results, CL-based recommenders inherit practical and conceptual issues. Negative sampling adds computational cost and design choices, and can interact unfavorably with long-tailed item distributions, potentially reinforcing popularity-driven concentration~\cite{ma2024negative,cai2024popularityawarealignmentcontrastmitigating,prakash2024evaluating}. In addition, CL performance can be sensitive to batch size and temperature, complicating deployment in large-scale or imbalanced regimes~\cite{DBLP:journals/corr/abs-2010-04592}. These limitations motivate non-contrastive alternatives that avoid explicit negatives and provide more direct control over representation geometry.

\paragraph{Non-Contrastive Learning and Redundancy Reduction.}
Non-contrastive self-supervised learning (NCL) learns invariant, non-collapsed representations without explicit negatives, using architectural asymmetries and regularizers. Representative approaches include BYOL~\cite{grill2020bootstrap}, SimSiam~\cite{chen2021exploring}, VICReg~\cite{bardes2022vicreg}, and Barlow Twins~\cite{zbontar2021barlow}, which differ in how they prevent collapse but often share a common mechanism: shaping representation geometry through variance reduction and feature decorrelation, including whitening-based objectives~\cite{ermolov2021whitening}. While these methods are well-studied in computer vision and have seen growing adoption in language domains~\cite{shiao2023linkpredictionnoncontrastivelearning}, their use in sequential recommendation remains relatively limited.
Most existing applications of NCL in recommendation follow an offline pretraining--fine-tuning pipeline. For example, CLUE~\cite{Cheng2021LearningTU} adapts BYOL-style training in collaborative filtering, and other works integrate self-supervision into matrix factorization~\cite{liu2021contrastiveselfsupervisedsequentialrecommendation}. Within sequential recommendation, NCL-SR~\cite{zeng2025a} extends non-contrastive learning but relies on external side information for view construction and does not study how feature decorrelation systematically affects recommendation behavior.

Our work departs from this line in two ways. First, we treat the redundancy-reduction objective as a \emph{feature-decorrelation regularizer} integrated \emph{directly into} next-item training, rather than as a standalone pretraining recipe. Specifically, we optimize $L_{\text{pred}}+\alpha L_{\text{decorr}}$, using Barlow Twins as a concrete instantiation of $L_{\text{decorr}}$. Second, we adopt target-item pairing (sequences sharing the same next item) to construct positive pairs, directly tying decorrelation to the recommendation objective without relying on synthetic corruptions. 
Beyond accuracy gains, we focus on \emph{mechanistic analysis}: how decorrelation reshapes embeddings and how this geometric change is related to the accuracy--diversity trade-off. 

\section{Decorrelation-Regularized Sequential Recommendation}
In this section, we introduce a \emph{decorrelation-regularized} training framework for sequential recommendation that enables us to study how redundancy reduction affects embedding geometry and the concentration of representations in the embedding space. The framework optimizes a standard next-item prediction objective augmented with a feature-decorrelation regularizer, $L_{\text{pred}}+\alpha L_{\text{decorr}}$, and uses \emph{target-item pairing} to form label-consistent positive pairs of user histories. Throughout the paper, we use \textbf{BT-SR} as a concrete instantiation where $L_{\text{decorr}}$ is the Barlow Twins loss; however, the same setup is compatible with other decorrelation objectives (e.g., VICReg~\cite{bardes2022vicreg} or whitening-based losses~\cite{huang2018decorrelated, ermolov2021whitening, hua2021feature}).



\subsection{Problem Setup and Preliminaries}
\label{sec:prelims}
We consider a standard sequential recommendation setting where the goal is to predict the next item a user interacts with, given their past behavior.
Let $\mathcal{U}$ denote the set of users and $\mathcal{I} = \{i_1, \dots, i_N\}$ the catalog of items. Each user $u \in \mathcal{U}$ has a chronological interaction sequence:
\begin{equation}
S_u = \bigl(i^u_{t_1}, i^u_{t_2}, \dots, i^u_{t_{|S_u|}}\bigr),
\end{equation}
where $t^u_k$ indexes interactions by timestamp. The goal of sequential recommendation is to learn a function that, given the $n$ most recent items from $S_u$, accurately predicts the next item the user will interact with. Formally, the model is trained to maximize the likelihood
$P_{\theta}\bigl(i^u_t \mid i^u_{t-n}, \dots, i^u_{t-1}\bigr)$,
where $\theta$ denotes model parameters.

In practice, this is implemented by encoding the user history at moment $t$ in the form of the item sequence \mbox{$s_u(t) = [i^u_{t-n}, \dots, i^u_{t-1}]$} into a representation $z_u(t) = f_\theta(s_u(t))$ using a Transformer encoder. The next-item prediction task is then formulated as a classification problem based on the relevance scores over the entire item catalog. The scoring function is typically defined in a matrix-factorization style as a dot product between an item embedding $e_i$ from the catalog and the current sequence state: $r_\theta(i, t) = e_i^\top z_u(t)$. We denote the classification loss as \(\mathcal{L}_{\text{pred}}\).

As baselines, we use SASRec variants trained with binary cross-entropy (BCE)~\cite{DBLP:journals/corr/abs-1808-09781}, full softmax cross-entropy (CE)~\cite{dross2023}, and scalable sampled softmax (SCE)~\cite{Mezentsev_2024}, described in detail in Section~\ref{subsec:baselines}.
While effective for optimizing next-item prediction accuracy, these objectives do not explicitly encourage structural alignment across semantically similar user sequences. In what follows, we introduce an augmentation scheme designed to reveal such alignment, followed by an auxiliary redundancy-reduction objective that strengthens the representational structure of the embedding space.

\subsection{Designing Supervised Augmentations}
\label{subsec:augs}
To induce semantically meaningful alignment between users, we design a supervised augmentation scheme guided by the next-item label. Inspired by prior work on label-guided contrastive learning~\cite{DBLP:journals/corr/abs-2110-05730}, we construct augmentations based on shared recent behavior. Specifically, given an anchor sequence $S_u = [i^u_{t-n}, \dots, i^u_{t}]$ ending with target item $i^u_t$, we uniformly sample another sequence $S_{u'}$ from the training set whose final item is also $i^u_t$. These two sequences, though originating from different users, reflect convergent behavioral patterns that are likely to lead to the same recommendation target.
We treat such sequence pairs $(S_u, S_{u'})$ as positive examples for the Barlow Twins loss. This encourages the model to produce similar embeddings for distinct consumption paths that converge to the same intent, while still allowing diversity across user histories.
This augmentation strategy is natural for the next-item prediction task and emphasizes behavioral convergence as a signal for alignment, allowing the model to generalize across different interaction paths while preserving user-specific information.

Crucially, we avoid applying random augmentations (e.g., masking, cropping, dropout-based perturbations), as we find that CL-style methods that heavily rely on such synthetic perturbations (e.g.,~\cite{xie2021contrastivelearningsequentialrecommendation, google25}) underperform our approach.
Moreover, they lack controllability over recommendation behavior that we aim to achieve.

\subsection{Feature-Decorrelation Regularization}
\label{sec:bt}

Building on the behavioral alignment provided by our augmentation scheme, we now define a redundancy-reduction objective that enhances the quality of sequence embeddings by promoting feature diversity and invariance. This objective, based on the Barlow Twins (BT) framework, is integrated as an auxiliary loss in our multi-task training setup.

Let \(Z^A\) and \(Z^B\) denote the original and augmented (by target-item anchoring) batches of sequence embeddings produced by the backbone network. We assume both views are \(\ell_2\)-normalized and mean-centered across the batch. We apply \(\ell_2\) normalization instead of the learned projection head originally used in the BT framework, as it is parameter-free and empirically more stable. The cross-correlation matrix \(\mathcal{C}\in\mathbb{R}^{D\times D}\) is computed as
\begin{align}\label{eq:1}
C_{ij}
&= \frac{1}{B}
\sum_{b=1}^B
\frac{Z^A_{b,i}\;Z^B_{b,j}}
{\sqrt{\sum_{b'=1}^B (Z^A_{b',i})^2}\;\sqrt{\sum_{b'=1}^B (Z^B_{b',j})^2}},
\end{align}
where \(b\) indexes samples in a batch of size \(B\), and \(i,j\) index embedding dimensions. Each \(C_{ij}\in[-1,1]\), with 1 indicating perfect correlation and -1 indicating perfect anti-correlation.

The corresponding BT loss is composed of two terms:
\begin{align}\label{eq:2}
\mathcal{L}_{BT}
&= \sum_{i=1}^D (1 - C_{ii})^2
\;+\;\lambda\sum_{i=1}^D\sum_{\substack{j=1\\j\neq i}}^D C_{ij}^2,
\end{align}
where \(\lambda\) trades off invariance against decorrelation. Driving the diagonal elements of $C$ toward $1$ enforces perturbation invariance, while pushing off-diagonals  toward $0$ reduces redundancy. 

We now define the complete loss function used to train BT-SR, combining next-item prediction with redundancy reduction in a single multi-task objective:
\begin{equation}
\mathcal{L}_{\text{total}} = \mathcal{L}_{\text{pred}} + \alpha \mathcal{L}_{\text{BT}},
\label{eq:total_loss}
\end{equation}
where \( \mathcal{L}_{\text{pred}} \) is the standard next-item prediction loss and \( \mathcal{L}_{\text{BT}} \) is the Barlow Twins redundancy-reduction loss. The hyperparameter \( \alpha \) modulates the influence of self-supervised regularization during training.
This formulation enables the model to learn embeddings that are both robust to behavioral variation and well-structured. As shown in Section~\ref{sec:results}, tuning \( \alpha \) allows practitioners to steer recommendation behavior in a controllable and interpretable way.


\subsection{Computational Complexity}
The computational overhead of the BT regularization is modest relative to the base SASRec model. SASRec itself has per-batch complexity $O(B \cdot L^2 \cdot D + B \cdot L \cdot D^2)$, where $B$ is the batch size, $L$ the sequence length, and $D$ the embedding dimension. Our BT-SR variant performs two forward passes (one per augmented view), doubling the cost of the backbone. Additionally, computing the cross-correlation matrix $\mathcal{C}$ adds $O(B \cdot D^2)$ operations per batch. Since $D$ is typically much smaller than $L^2$ or $L \cdot D$ for practical sequence lengths, the overall asymptotic complexity remains \mbox{$O(B \cdot L^2 \cdot D + B \cdot L \cdot D^2)$}, matching that of standard SASRec. Thus, the redundancy-reduction objective introduces only a small constant-factor training-time slowdown in practice.

\subsection{Feature Decorrelation and Popularity Bias}
\label{sec:decorrelation_popbias}

We provide a geometric interpretation of how adding a feature-decorrelation term modulates the relationship between embedding geometry and popularity effects in dot-product recommenders.
\paragraph{Shared Low-Rank Structure View on Item Popularity.}
Assume items are ranked by the score ${s(z_u,i)=z_u^\top e_i}$, where ${z_u\in\mathbb{R}^d}$ is the user-history representation and ${e_i\in\mathbb{R}^d}$ is the item embedding.
The training dynamics of sequential recommenders induce a shared low-rank structure in the user representation space: items that appear often as targets undergo more frequent updates, causing their embeddings to converge toward directions that are common across user histories.
To formalize the notion of ``common directions'', write ${z_u=\mu+\varepsilon}$ with ${\mu=\mathbb{E}[z_u]}$ and ${\mathbb{E}[\varepsilon]=0}$, and let ${\Sigma_{z_u}=\mathrm{Cov}(z_u)}$.
If $\mu\neq 0$, we may decompose any item embedding $e_i$ as
\begin{equation}
e_i = a_i\frac{\mu}{\|\mu\|}+r_i,\qquad r_i\perp \mu,
\end{equation}
where $a_i$ is the projection of $e_i$ onto the direction of $\mu$. The score then becomes
\begin{equation}
s(z_u,i)=z_u^\top e_i
= a_i\frac{z_u^\top \mu}{\|\mu\|}+z_u^\top r_i.
\end{equation}
The first term acts as an item-specific but user-agnostic component shared across users through the scalar $z_u^\top \mu$. If frequent items have a larger projection onto this shared direction (larger $a_i$), they receive a systematic score advantage across many users.

More generally, even when ${\mu\approx 0}$, shared structure can be captured by the dominant eigendirections of $\Sigma_{z_u}$. Denote the top-$m$ eigenvectors of $\Sigma_{z_u}$ as ${Q_m=[q_1,\dots,q_m]\in\mathbb{R}^{d\times m}}$. Items with large energy in this shared subspace,
\begin{equation}
g_i^{(m)} \;=\; \frac{\|Q_m^\top e_i\|^2}{\|e_i\|^2}
\;=\; \sum_{k=1}^m \frac{(q_k^\top e_i)^2}{\|e_i\|^2},
\end{equation}
are particularly sensitive to global factors common across users, which increases their propensity to appear in top-$K$ lists for many users and thus exacerbates popularity-driven concentration. In what follows we fix $m=2$ and omit the superscript for clarity.

\paragraph{Measuring Propensity to Popularity-Driven Concentration.}
Let $\bar g_b$ denote the average of $g_i$ over items in popularity bucket $b$, where $b=1$ is the head (most popular) and larger $b$ correspond to progressively less popular items (the tail). Since absolute alignment levels can differ across methods due to differences in embedding geometry, we normalize each curve by the head bucket: $\tilde g_b=\bar g_b/\bar g_1$, allowing us to compare relative head-to-tail changes rather than raw magnitudes. We then summarize the head-to-tail change as an \emph{Alignment Concentration} score
\begin{equation}
\mathrm{AC} \;=\; \sum_{b>1}\big( 1 - \tilde g_b\big),
\end{equation}
where higher values correspond to stronger concentration of shared-subspace alignment in the head bucket. We use this score as a proxy for a model's propensity to popularity-driven concentration in Section~\ref{sec:ac} (see Figure~\ref{figure:ml_alignment}).


\paragraph{Effect of Decorrelation.}
A feature-decorrelation objective counteracts the above mechanism by suppressing shared directions and promoting approximately zero-mean, isotropic representations. In particular, if decorrelation encourages ${\mathbb{E}[z_u]\approx 0}$ and ${\Sigma_{z_u}\approx I}$, then
\[
\mathbb{E}_{z_u}[s(z_u,i)] = \mathbb{E}[z_u]^\top e_i \approx 0,
\quad
\mathbb{V}_{z_u}[s(z_u,i)] = e_i^\top \Sigma_{z_u}\,e_i \approx \|e_i\|^2,
\]
so neither a global mean direction nor a small set of dominant principal directions can provide a broad advantage across users. Consequently, recommendation scores rely less on shared low-rank structure (i.e., high $g_i$) and more on user-conditional residual components, which may reduce popularity-driven concentration and improve long-tail exposure.


\section{Experimental Setup}

\subsection{Datasets} 
Experiments are conducted on five public datasets: Behance~\cite{behance}, Kindle Store~\cite{ni-etal-2019-justifying}, Yelp~\cite{asghar2016yelp}, Gowalla~\cite{gowalla}, and Movielens-1M (ML-1M)~\cite{movielens2015}. Users and items with fewer than five interactions are filtered out. We follow the global timepoint split (GTS) experimental protocol~\cite{gusak2025time}, which ensures controlled comparison and prevents temporal leakage. A global timestamp at the 0.95 quantile defines the boundary between training and test data. For each test user, we apply a GTS-respecting leave-one-out protocol, using the latest interaction for testing and the second-to-last for validation. The remaining items between GTS timepoint and held-out items are appended to test users histories during inference.

All experiments are conducted on a single NVIDIA A100 GPU. 
The code, preprocessing scripts and additional ablation studies and experiments are publicly available\footnote{\url{https://github.com/Veronika-Ivanova/barlow_twins_sasrec}\label{fn:repo}}.

\subsection{Metrics} 
Following best practices~\cite{ca2020, dallmann2021, krichene2020}, we evaluate models using unsampled top-$K$ ranking metrics computed over the full item catalog. 
We report Normalized Discounted Cumulative Gain (ndcg@K) and Hit Rate (hr@K) for \mbox{$K = 1, 5, 10$}, averaged across all test users. 
To assess recommendation diversity, we report item coverage (cov@K), defined as the fraction of unique items appearing in top-$K$ recommendations across all users. We further apply a frequency-bucket analysis in Section~\ref{sec:analysis}, splitting the item catalog into buckets of varying interaction frequency with roughly equal interaction counts, which provides a more detailed view of recommendation quality across different item groups.

\subsection{Hyperparameters} 
All models are implemented in PyTorch and optimized with Adam using a learning rate of 0.001. 
The maximum sequence length is fixed to 50, truncating longer histories to the most recent interactions. 
L2 weight decay is applied for regularization, with the coefficient tuned on the validation set (typically \(10^{-4}\)–\(10^{-5}\)).
First, we tune the Barlow Twins hyperparameters $\alpha$ and $\lambda$ jointly with all other hyperparameters. Subsequently, in an ablation study, we perform a dedicated grid search over $\alpha$ and $\lambda$ using values from ${0.05, 0.10, \dots, 0.50}$, selecting the best based on validation performance.
The auxiliary loss \( \mathcal{L}_{BT} \) is applied to different SASRec variants trained with \( \mathcal{L}_{\text{BCE}} \)~\cite{DBLP:journals/corr/abs-1808-09781}, \( \mathcal{L}_{\text{CE}} \)~\cite{dross2023}, and \( \mathcal{L}_{\text{SCE}} \)~\cite{Mezentsev_2024}. 
We report the best-performing model for each dataset. 
To ensure statistical significance, we report the mean and standard deviation over five independent runs. 
Significance is verified using paired t-tests ($p < 0.05$) against the second-best baseline, with statistically significant improvements marked by an asterisk (*). 
The low observed variance demonstrates the stability of our method.
\begin{figure}[!t]
    \centering

    \includegraphics[width=0.5\linewidth]{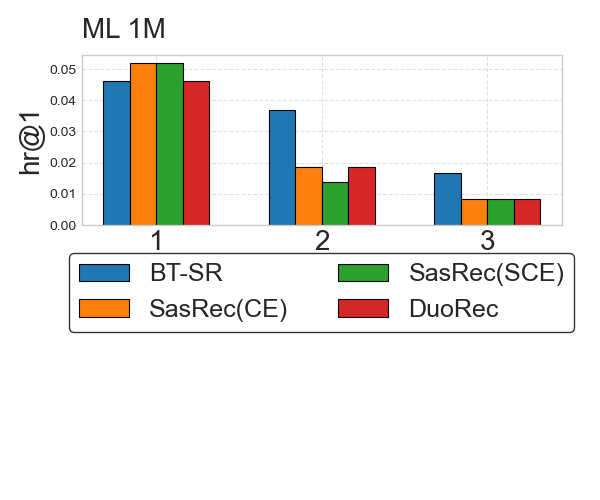}
    \includegraphics[width=0.49\linewidth]{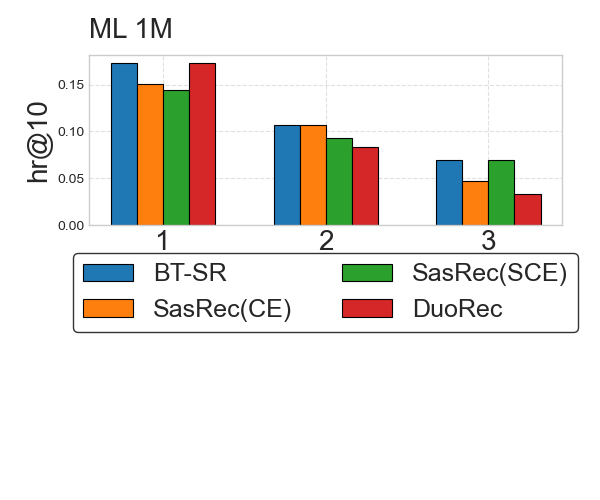}
    \\[-30pt]
    
    \includegraphics[width=0.5\linewidth]{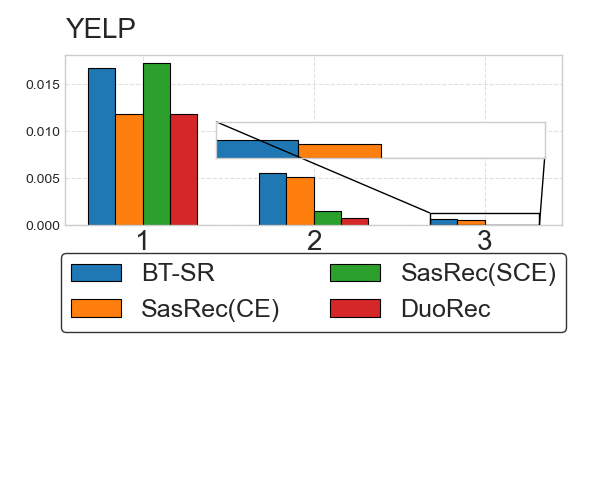}
    \includegraphics[width=0.49\linewidth]{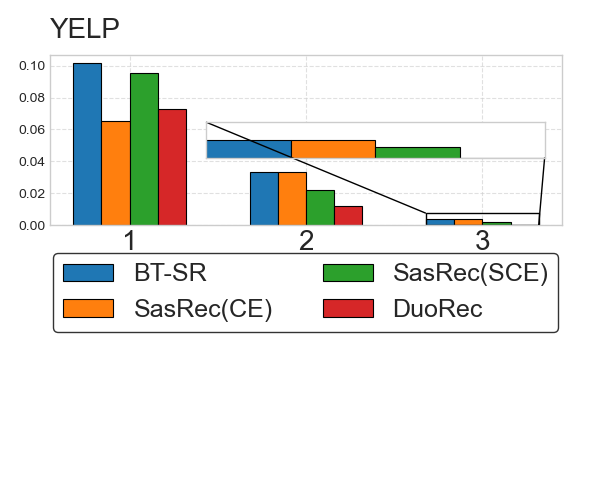}
    \\[-30pt]
    
    \includegraphics[width=0.5\linewidth]{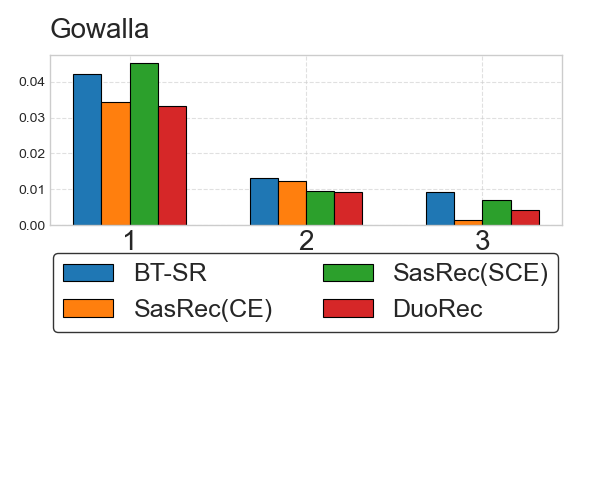}
    \includegraphics[width=0.49\linewidth]{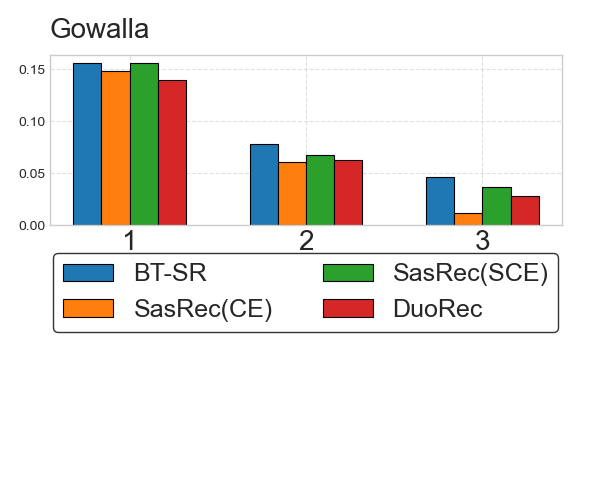}
    \\[-32pt]
    
    \caption{HR@1 (left) and HR@10 (right) metrics for the three item-popularity buckets across three datasets.}
    \label{figure:pop_buckets}
    \Description[Grouped bar charts comparing HR@1 (left) and HR@10 (right) for baseline models: SASRec(CE), SASRec(SCE), DuoRec, and our BT-SR across head, mid, and tail item-popularity buckets on three datasets.]{Figure 1 contains two grouped bar-chart panels. The left panel shows HR@1 and the right panel shows HR@10, each broken down by three item-popularity buckets (head/most popular, mid, and tail/least popular) for three datasets. Bars compare the baseline models: SASRec(CE), SASRec(SCE) and DuoRec, against the proposed BT-SR model. Across datasets, BT-SR is slightly below the baseline in the head bucket for HR@1 but clearly higher for the mid and tail buckets, indicating improved personalization on less popular items. For HR@10, BT-SR matches or exceeds the baseline even in the head bucket, showing that its overall ranking accuracy is not sacrificed while long-tail items are surfaced more often.}
\end{figure}

\begin{table}[b]
\caption{NDCG@10 performance comparison of three SASRec variants (BCE loss, CE loss, SCE loss) with and without the Barlow Twins regularization term. Results highlight consistent improvements from Barlow Twins across all scenarios.}
\label{table:utility_backbones}
    {\footnotesize
        \begin{tabular}{rcc|cc|cc}
        \toprule
         & BCE & BCE + BT & CE & CE + BT & SCE & SCE + BT \\
        \midrule
        ML-1M & 0.0443 & \textbf{0.0490} & 0.0494 & \textbf{0.0583} & 0.0503 & \textbf{0.0562} \\
        YELP & 0.0146 & \textbf{0.0149} & 0.0127 & \textbf{0.0150} & 0.0137 & \textbf{0.0141} \\
        Gowalla & 0.0419 & \textbf{0.0434} & 0.0364 & \textbf{0.0481} & 0.0457 & \textbf{0.0483} \\
        Beauty & 0.0434 & \textbf{0.0440} & 0.0524 & \textbf{0.0563} & 0.0543 & \textbf{0.0549} \\
        Kindle Store & 0.0705 & \textbf{0.0733} & 0.0700 & \textbf{0.0746} & 0.0765 & \textbf{0.0796} \\
        \bottomrule
        \end{tabular}
    }
\end{table}

\begin{table*}[!t]
    \caption{Performance comparison across all datasets. The best result in each row is highlighted in bold, the second-best result is underlined. All metrics are averaged over 5 independent runs,  statistically significant improvements are marked by (*).}
    {\footnotesize
    \begin{tabular}{rlcccccccccl}
    \toprule
    Dataset & Metric & SasRec(BCE) & SasRec(CE) & SasRec(SCE) & CL4SRec & DuoRec & EC4SRec & BT-PT & \textbf{BT-SR}(Ours) & Improve \\
    \midrule
    \multirow{9}{*}{ML-1M}
 & hr@1    & 0.0160                   & \underline{0.0213}       & 0.0200                   & 0.0182                   & 0.0200                   & \underline{0.0213}                   & 0.0200                   & \textbf{0.0240* $ \pm $ 0.0048} & +12.7\% \\
 & hr@5    & 0.0493                   & \underline{0.0600}       & 0.0587                   & 0.0476                   & 0.0480                   & 0.0534                   & 0.0573                   & \textbf{0.0639* $ \pm $ 0.0024} & +6.5\% \\
 & hr@10   & 0.0893                   & 0.0847                   & \underline{0.0933}       & 0.0834                   & 0.0800                   & 0.0880                   & 0.0893                   & \textbf{0.0945* $ \pm $ 0.0080} & +1.3\% \\
  \arrayrulecolor{gray!30}\cmidrule(lr){2-11}\arrayrulecolor{black}
 & ndcg@5  & 0.0319                   & \underline{0.0404}       & 0.0393                   & 0.0342                   & 0.0353                   & 0.0364                   & 0.0383                   & \textbf{0.0436* $ \pm $ 0.0020} & +7.9\% \\ 
 & ndcg@10 & 0.0443                   & 0.0494                   & \underline{0.0503}       & 0.0464                   & 0.0456                   & 0.0487                   & 0.0486                   & \textbf{0.0534* $ \pm $ 0.0037} & +6.2\% \\
\arrayrulecolor{gray!30}\cmidrule(lr){2-11}\arrayrulecolor{black}
 & cov@1   & 0.0320                   & \underline{0.0887}       & \textbf{0.0961}          & 0.0335                   & 0.0361                   & 0.0387                   & 0.0846                   & 0.0816* $ \pm $ 0.0035          & -15.1\% \\
 & cov@5   & 0.1473                   & 0.2265       & \textbf{0.2568}          & 0.0969                   & 0.0936                   & 0.1002                   & \underline{0.2267}        & 0.2223* $ \pm $ 0.0083          & -13.4\% \\
 & cov@10  & 0.2434                   & 0.3185                   & \textbf{0.3645}          & 0.1329                   & 0.1459                   & 0.1591                   & \underline{0.3193}        & 0.3189* $ \pm $ 0.0123          & -12.5\% \\

    \midrule
    \multirow{9}{*}{YELP}
 & hr@1   & \underline{0.0045} & 0.0043 & 0.0040 & 0.0028 & 0.0027 & 0.0028 & 0.0030 & \textbf{0.0049* $\pm$ 0.0005}  & +8.9\% \\
 & hr@5   & \underline{0.0162} & 0.0137 & 0.0152 & 0.0087 & 0.0094 & 0.0090 & 0.0121 & \textbf{0.0183* $\pm$ 0.0008}  & +13.0\% \\
 & hr@10  & \textbf{0.0292} & 0.0257 & 0.0277 & 0.0192 & 0.0188 & 0.0199 & 0.0244 & \underline{0.0288 $\pm$ 0.0010}  & –1.4\% \\
\arrayrulecolor{gray!30}\cmidrule(lr){2-11}\arrayrulecolor{black}
 & ndcg@5 & \underline{0.0104} & 0.0088 & 0.0097 & 0.0060 & 0.0061 & 0.0062 & 0.0074 & \textbf{0.0117* $\pm$ 0.0006}  & +12.5\% \\
 & ndcg@10& \underline{0.0146} & 0.0127 & 0.0137 & 0.0085 & 0.0091 & 0.0092 & 0.0114 & \textbf{0.0150* $\pm$ 0.0006}  & +2.7\%  \\
\arrayrulecolor{gray!30}\cmidrule(lr){2-11}\arrayrulecolor{black}
 & cov@1  & 0.0103 & \textbf{0.0233} & 0.0135 & 0.0057 & 0.0059 & 0.0059 & 0.0159 & \underline{0.0218 $\pm$ 0.0034}  & –6.4\%  \\
 & cov@5  & 0.0333 & \underline{0.0621} & 0.0413 & 0.0170 & 0.0161 & 0.0171 & 0.0444 & \textbf{0.0636* $\pm$ 0.0089}  & +2.4\%  \\
 & cov@10 & 0.0541 & \underline{0.0911} & 0.0661 & 0.0276 & 0.0255 & 0.0275 & 0.0678 & \textbf{0.0998* $\pm$ 0.0128}  & +9.5\% \\

    \midrule
    \multirow{9}{*}{Gowalla}
 & hr@1   & 0.0173 & 0.0143 & \underline{0.0196} & 0.0157 & 0.0144 & 0.0142 & 0.0184 & \textbf{0.0205* $\pm$ 0.0010} & +4.6\%  \\
 & hr@5   & 0.0506 & 0.0431 & \underline{0.0548} & 0.0433 & 0.0473 & 0.0440 & 0.0502 & \textbf{0.0592* $\pm$ 0.0010}  & +8.0\%  \\
 & hr@10  & 0.0756 & 0.0660 & \underline{0.0811} & 0.0694 & 0.0712 & 0.0757 & 0.0728 & \textbf{0.0851* $\pm$ 0.0020}  & +4.9\%  \\
\arrayrulecolor{gray!30}\cmidrule(lr){2-11}\arrayrulecolor{black}
 & ndcg@5 & 0.0339 & 0.0289 & \underline{0.0372} & 0.0318 & 0.0310 & 0.0331 & 0.0344 & \textbf{0.0400* $\pm$ 0.0009}  & +7.5\%  \\
 & ndcg@10& 0.0419 & 0.0364 & \underline{0.0457} & 0.0411 & 0.0387 & 0.0397 & 0.0417 & \textbf{0.0483* $\pm$ 0.0013}  & +5.7\%  \\
\arrayrulecolor{gray!30}\cmidrule(lr){2-11}\arrayrulecolor{black}
 & cov@1  & 0.0230 & 0.0228 & 0.0214 & 0.0252 & 0.0255 & 0.0270 & \textbf{0.0368} & \underline{0.0321* $\pm$ 0.0039}  & -12.8\% \\
 & cov@5  & 0.0892 & 0.0725 & 0.0921 & 0.0950 & 0.0939 & 0.0998 & \underline{0.1269} & \textbf{0.1280* $\pm$ 0.0140}  & +7.3\% \\
 & cov@10 & 0.1557 & 0.1153 & 0.1625 & 0.1538 & 0.1595 & 0.1650 & \underline{0.2037} & \textbf{0.2187* $\pm$ 0.0237}  & +7.6\% \\

    \midrule
    \multirow{9}{*}{Beauty}
 & hr@1    & 0.0179                   & 0.0269                   & \underline{0.0305}       & 0.0292                   & \underline{0.0305}       & 0.0303                   & 0.0278                   & \textbf{0.0326* $ \pm $ 0.0013} & +6.9\% \\
 & hr@5    & 0.0538                   & 0.0591       & 0.0582                   & 0.0555                   & 0.0573                   & 0.0574                   & \underline{0.0601}           & \textbf{0.0606* $ \pm $ 0.0025} & +2.5\% \\
 & hr@10   & 0.0789                   & 0.0896                   & 0.0905       & 0.0830                   & 0.0860                   & 0.0815                   & \underline{0.0928}           & \textbf{0.0937* $ \pm $ 0.0019} & +3.5\% \\
\arrayrulecolor{gray!30}\cmidrule(lr){2-11}\arrayrulecolor{black}
 & ndcg@5  & 0.0355                   & 0.0425                   & 0.0442                   & 0.0412                   & 0.0436                   & \underline{0.0455}       & 0.0441                   & \textbf{0.0463* $ \pm $ 0.0002} & +1.8\% \\
 & ndcg@10 & 0.0434                   & 0.0524                   & 0.0543                   & 0.0530                   & 0.0527                   & 0.0537                   & \underline{0.0551}        & \textbf{0.0569* $ \pm $ 0.0001} & +4.8\% \\
\arrayrulecolor{gray!30}\cmidrule(lr){2-11}\arrayrulecolor{black}
 & cov@1   & 0.0620                   & 0.0546                   & 0.0499                   & 0.0603                   & 0.0607                   & \textbf{0.0679}          & \underline{0.0634}        & 0.0491* $ \pm $ 0.0078          & -27.7\% \\
 & cov@5   & \underline{0.2200}       & 0.1782                   & 0.1760                   & \textbf{0.2212}          & 0.2100                   & 0.2103                   & 0.2130                   & 0.1532* $ \pm $ 0.0357          & -29.2\% \\
 & cov@10  & \textbf{0.3448}          & 0.2752                   & 0.2766                   & \underline{0.3362}       & 0.3234                   & 0.3236                   & 0.3272                   & 0.3234 $ \pm $ 0.0612          & -6.2\% \\

    \midrule
    \multirow{9}{*}{Kindle Store}
 & hr@1    & 0.0451                   & 0.0471                   & \underline{0.0533}       & 0.0469                   & 0.0471                   & 0.0505                   & 0.0502                   & \textbf{0.0564* $ \pm $ 0.0007} & +5.8\% \\
 & hr@5    & 0.0832                   & 0.0813                   & \underline{0.0899}       & 0.0813                   & 0.0762                   & 0.0785                   & 0.0820                   & \textbf{0.0903* $ \pm $ 0.0018} & +0.4\% \\
 & hr@10   & 0.0990                   & 0.0965                   & \underline{0.1022}       & 0.0910                   & 0.0897                   & 0.0955                   & 0.0953                   & \textbf{0.1058* $ \pm $ 0.0012} & +3.5\% \\
\arrayrulecolor{gray!30}\cmidrule(lr){2-11}\arrayrulecolor{black}
 & ndcg@5  & 0.0655                   & 0.0652                   & \underline{0.0726}       & 0.0635                   & 0.0625                   & 0.0652                   & 0.0678                   & \textbf{0.0747* $ \pm $ 0.0009} & +2.9\% \\
 & ndcg@10 & 0.0705                   & 0.0700                   & \underline{0.0765}       & 0.0662                   & 0.0668                   & 0.0672                   & 0.0721                   & \textbf{0.0796* $ \pm $ 0.0005} & +4.1\% \\
\arrayrulecolor{gray!30}\cmidrule(lr){2-11}\arrayrulecolor{black}
 & cov@1   & 0.0396                   & 0.0337                   & \underline{0.0409}       & 0.0363                   & 0.0356                   & 0.0369                   & 0.0387                   & \textbf{0.0440* $ \pm $ 0.0005} & +7.6\% \\
 & cov@5   & 0.1278                   & 0.1076                   & \underline{0.1492}       & 0.1077                   & 0.1123                   & 0.1082                   & 0.1304                   & \textbf{0.1665* $ \pm $ 0.0019} & +11.6\% \\
 & cov@10  & 0.1920                   & 0.1665                   & \underline{0.2346}       & 0.1598                   & 0.1719                   & 0.1703                   & 0.2016                   & \textbf{0.2673* $ \pm $ 0.0021} & +13.9\% \\
    \bottomrule
\end{tabular}
    }

    \label{table:utility_metrics}
\end{table*}

\subsection{Baselines}
\label{subsec:baselines}

As baselines, we adopt \textbf{SASRec} variants trained with: binary cross-entropy (BCE)~\cite{DBLP:journals/corr/abs-1808-09781}, standard cross-entropy (CE)~\cite{dross2023}, and scalable cross-entropy (SCE)~\cite{Mezentsev_2024}.
For contrastive-learning baselines, we include \textbf{CL4SRec}~\cite{xie2021contrastivelearningsequentialrecommendation}, \textbf{DuoRec}~\cite{DBLP:journals/corr/abs-2110-05730}, and \textbf{EC4SRec}~\cite{wang2022explanationguidedcontrastivelearning}.
\textbf{BT-PT} denotes a pretraining-based baseline inspired by~\cite{google25}; to ensure fair comparison, we use a \textbf{SASRec} backbone trained in two stages with Barlow Twins self-supervised pretraining followed by supervised next-item prediction. We set the augmentation probability to $p=0.2$, as in the original paper~\cite{google25}, and tune the redundancy-reduction coefficient $\lambda$ for this method.

While recent LLM- and LoRA-based recommenders perform well in text-rich or cold-start settings, they underperform in ID-based domains and incur substantially higher inference costs. 
Moreover, prior studies~\cite{10.1145/3726302.3730178, DBLP:journals/corr/abs-2503-05493} show that large language models often memorize public recommendation datasets, raising concerns of target leakage rather than genuine generalization. 
For these reasons, we exclude LLM-based baselines from our evaluation.

\section{Results}
\label{sec:results}

We first evaluate the impact of adding the Barlow Twins term to three SASRec objectives (\(\mathcal{L}_{BCE}\), \(\mathcal{L}_{CE}\), \(\mathcal{L}_{SCE}\)). As shown in Table~\ref{table:utility_backbones}, the Barlow Twins term consistently improves performance across all base losses in the multi-task setup, demonstrating the universality of our approach.  

Next, we select the best-performing objective for each dataset. \(\mathcal{L}_{SCE}\) achieves the strongest results on Gowalla and Kindle Store, while \(\mathcal{L}_{CE}\) performs better on the remaining datasets. Using these selected models, we compare against strong baselines on the holdout test set. Table~\ref{table:utility_metrics} reports our best variant (BT-SR), which consistently outperforms all baselines in both utility and coverage across most datasets. To further investigate the role of the Barlow Twins loss in representation learning, we provide a detailed analysis in the following sections.

\subsection{Accuracy Analysis Across Item Frequency Cohorts}
\label{sec:analysis}

While BT-SR underperforms the best baseline in terms of coverage on ML-1M and Amazon Beauty, it achieves notably stronger coverage on the remaining three datasets. To unpack this behavior, we order items in each catalog from most to least popular and split them into three popularity-based buckets so that the first bucket contains more popular items and the last bucket consists of long-tailed ones. For balanced evaluation, we ensure each bucket covers roughly one-third of total interactions. Figure~\ref{figure:pop_buckets} shows HR@1 and HR@10 for each bucket.

Surprisingly, BT-SR underperforms slightly at HR@1 in the top-popularity bucket, yet it markedly outperforms all baselines on the mid- and low-popularity buckets --- evidence of its enhanced personalization. Moreover, BT-SR also leads at HR@10 even for the most popular items, indicating that \emph{it can elevate niche items into top-rank positions without sacrificing performance on blockbusters}. This pattern highlights how the Barlow Twins loss both prevents representation collapse and \emph{promotes more uniform treatment of items across frequency groups}.

Further, in Section~\ref{sec:ablation} we demonstrate that BT-SR can be tuned for industrial applications via an optional hyperparameter that explicitly balances HR@1 against HR@10, allowing practitioners to prioritize the metric that best suits their use case.






\begin{figure}[!t]
    \centering

    \includegraphics[width=1\linewidth]{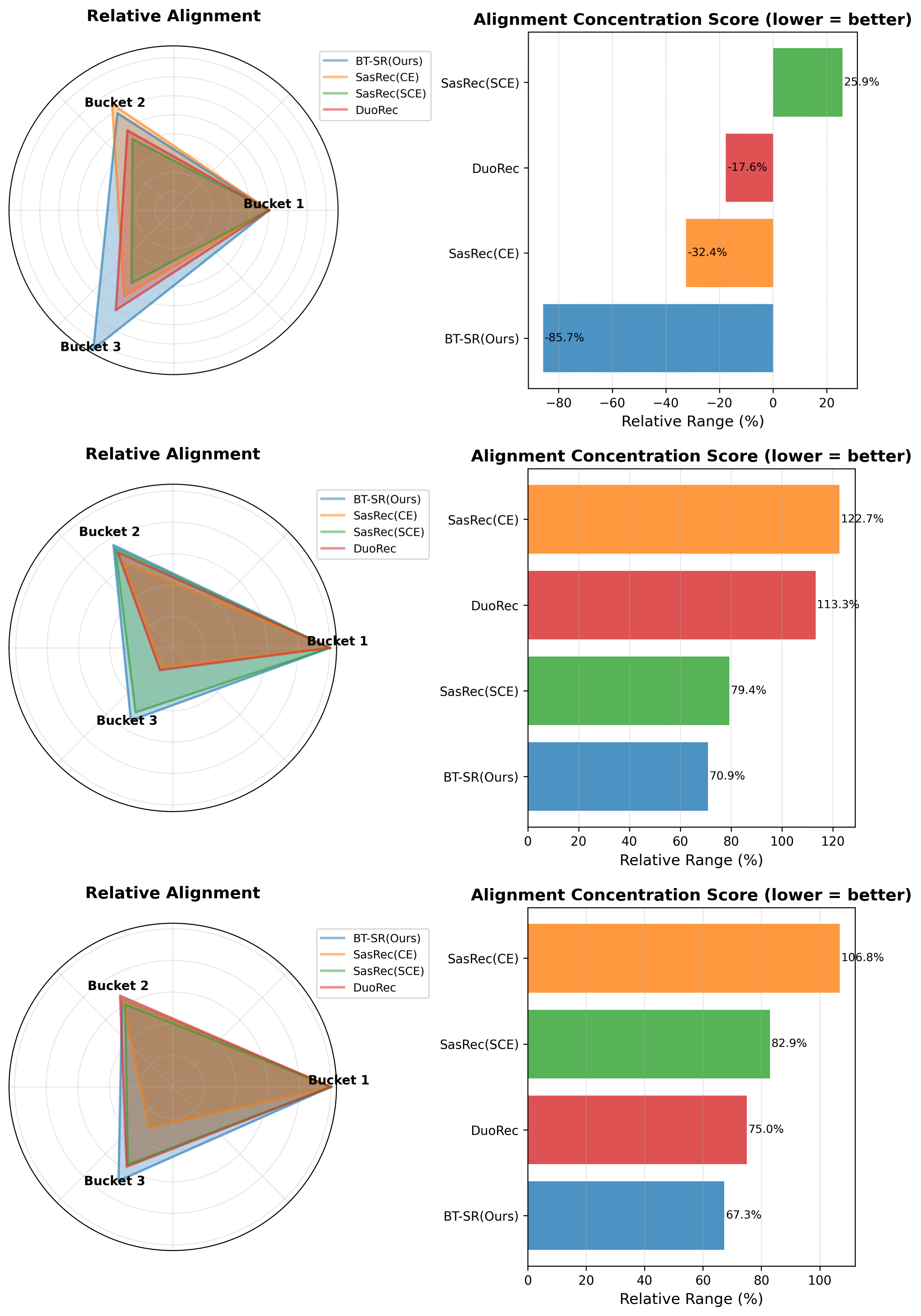}

    \caption{
    Alignment Concentration on Movielens-1M, Yelp and Gowalla datasets. (left) We plot each model in relative alignment, $\tilde g_b$, coordinates. (right) The AC score quantifies, across all tail buckets, how much the alignment of less popular items falls short of the alignment of head-bucket items.}
    \label{figure:ml_alignment}
    \Description[Left: radar plots of relative head-normalized alignment across three popularity buckets for BT-SR, SASRec(CE), SASRec(SCE) and DuoRec models on MovieLens-1M, Yelp, and Gowalla. Right: bar chart of the Alignment Concentration (AC) score per model per dataset, with BT-SR showing the lowest values.]{Figure 2 has two parts. The left part shows three radar (spider) plots, one per dataset (MovieLens-1M, Yelp, Gowalla), where each axis corresponds to one of three item-popularity buckets and each colored line represents a different model (SASRec-CE, SASRec-SCE, DuoRec, BT-SR, etc.). The head bucket is normalized to 1, and the plotted values ($g_b$) show how strongly items in the mid and tail buckets align with the shared low-rank user-representation subspace relative to the head. BT-SR's line is visibly flatter or extends beyond 1 in the outer buckets compared to competing models, meaning it is less head-concentrated. The right part is a grouped bar chart showing the scalar Alignment Concentration (AC) score for each model on each dataset; lower values indicate weaker head-item dominance. BT-SR consistently has the lowest (or most negative) AC score across all three datasets, confirming it reduces reliance on shared low-rank directions that favor popular items.}
\end{figure}

\subsection{Alignment Concentration Scores}
\label{sec:ac}

\begin{figure}[!t]
    \centering

    \includegraphics[width=1\linewidth]{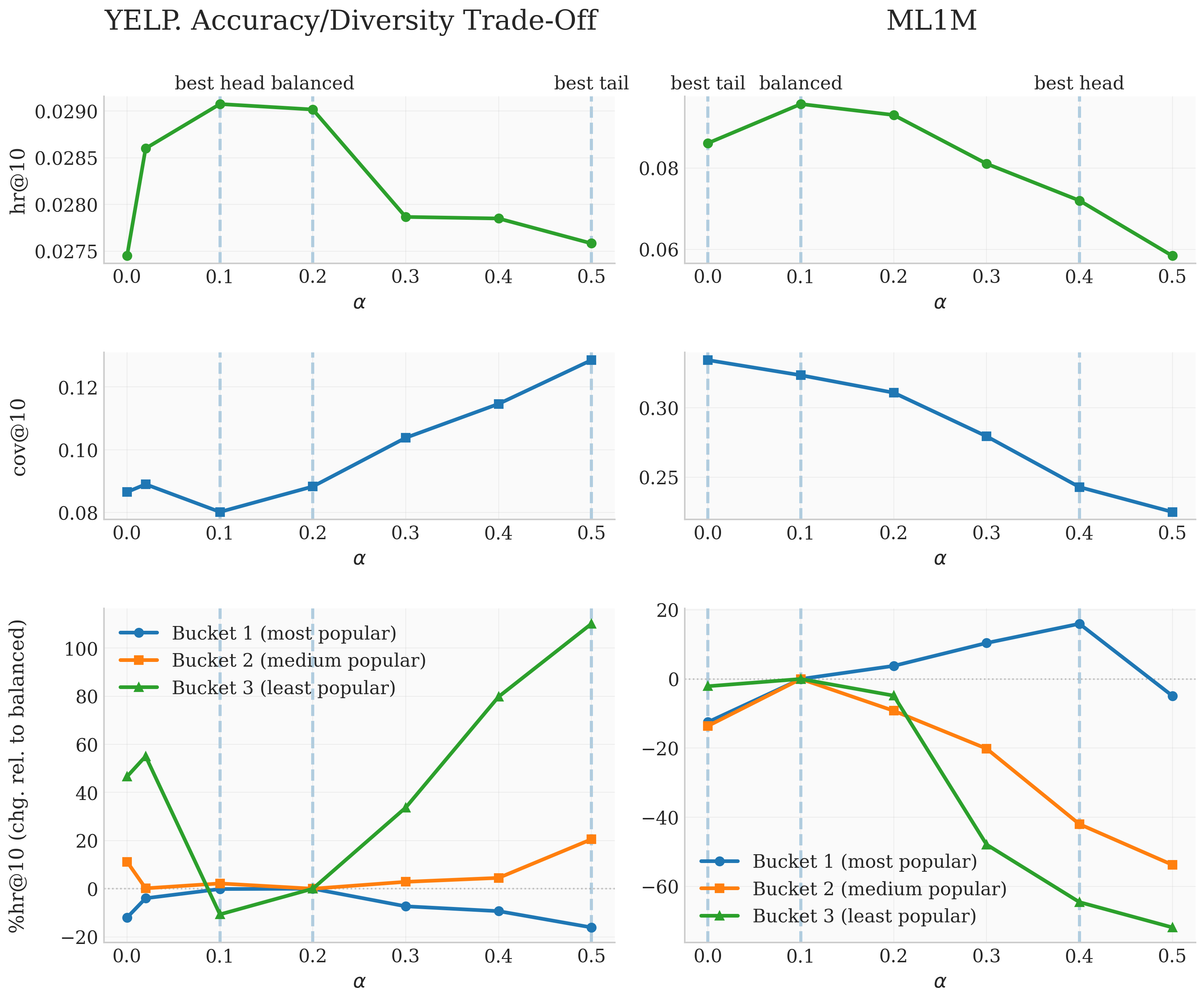}

    \caption{Effect of the trade-off parameter $\alpha$ on accuracy (HR@10) and coverage (COV@10) for YELP and MovieLens-1M.
Results are reported over the full catalog, along with relative changes of HR@10 across three popularity buckets (w.r.t. balance point).
Varying $\alpha$ induce head-oriented, balanced, and tail-oriented recommendation regimes. }
    \label{figure:ml_abol}
    \Description[Line plots showing HR@10 and COV@10 as functions of the decorrelation weight $\alpha$ for Yelp and MovieLens-1M, plus relative HR@10 change across three popularity buckets.]{Figure 3 presents line charts with the decorrelation strength $\alpha$ (ranging from 0 to 0.5) on the x-axis, shown separately for Yelp and MovieLens-1M. For each dataset, one curve shows HR@10 and another shows COV@10 (item coverage), both computed over the full catalog as $\alpha$ increases. An additional set of curves shows the relative change in HR@10 for the head, mid, and tail popularity buckets relative to a chosen 'balance point' value of $\alpha$.

For Yelp, increasing $\alpha$ beyond the balance point produces a steady rise in COV@10, while HR@10 declines only moderately. The bucket-wise curves for Yelp show substantial HR@10 gains for the least-frequent (tail) items and only mild losses for the most-frequent (head) items, indicating that stronger decorrelation shifts recommendation mass toward long-tail items while preserving competitive overall accuracy.

For MovieLens-1M, the pattern differs: the best COV@10 is achieved at small $\alpha$ values, while larger $\alpha$ reduces both HR@10 and COV@10. The bucket-wise breakdown for MovieLens-1M shows that this degradation is concentrated in the medium- and low-frequency groups, whereas HR@10 for the most frequent items actually improves as $\alpha$ grows — the opposite direction from Yelp's trend.

Together, the two datasets illustrate that $\alpha$ acts as a tunable knob that reshapes the head-tail allocation of recommendation accuracy, but the direction and magnitude of this trade-off is dataset-dependent, likely reflecting differences in interaction sparsity and timestamp/popularity structure between Yelp and MovieLens-1M.}
\end{figure}
\paragraph{Head--tail alignment profiles and concentration}
Figure~\ref{figure:ml_alignment} reports the head-normalized alignment profiles $\tilde g_b=\bar g_b/\bar g_1$ across three popularity buckets (left) together with the corresponding Alignment Concentration score $\mathrm{AC}=\sum_{b>1}(1-\tilde g_b)$ (right; lower is better, as it indicates less head concentration). 
Intuitively, $\tilde g_b$ captures how strongly items in bucket $b$ align with the shared low-rank user subspace relative to the head bucket, and $\mathrm{AC}$ summarizes the extent to which this alignment is concentrated. Larger positive $\mathrm{AC}$ indicates a pronounced head--tail gap (strong head concentration), values near zero indicate a flat profile, and negative $\mathrm{AC}$ indicates an inverted pattern where some non-head buckets align \emph{more} strongly than the head.

\paragraph{Consistent reduction of head concentration under BT-SR}
Across all three datasets, BT-SR yields the lowest AC score, indicating that the alignment of item embeddings to the shared user subspace is less concentrated in the head bucket under BT-SR than under other models. This effect is visible in the radar plots in Figure~\ref{figure:ml_alignment}: compared to SASRec (CE/SCE) and DuoRec, BT-SR produces a flatter (or even inverted) head-to-tail profile --- that is, $\tilde g_2$ and $\tilde g_3$ remain closer to (or exceed) $1$ rather than dropping sharply below the head baseline. In the language of Section~\ref{sec:decorrelation_popbias}, BT-SR exhibits a weaker tendency for popular items to dominate rankings via shared low-rank directions in the user-history representation space.

\paragraph{Dataset-dependent behavior and interpretation}
We also observe that the absolute magnitude and even the sign of $\mathrm{AC}$ can vary across datasets and objectives. In particular, negative $\mathrm{AC}$ implies that shared-subspace alignment is not head-concentrated under that method, suggesting that popularity-driven concentration (when present) may be dominated by other factors (e.g., embedding norms or exposure dynamics) rather than alignment to shared directions. Nevertheless, the relative ordering remains stable: BT-SR consistently minimizes $\mathrm{AC}$, supporting our central claim that feature decorrelation reduces reliance on shared low-rank structure that amplifies head-item dominance.
We find that lower AC is associated with reduced head share in top-K recommendations and improved long-tail exposure, suggesting that the proposed metric captures a meaningful driver of head-to-tail recommendation distribution.


\begin{table}[!t]
\caption{Ablation on item-wise sequence augmentations (as in CL4SRec) and replacing the SSL objective with SimSiam.}
{\footnotesize
\begin{tabular}{rlcccc}
\toprule
Dataset & Metric
& \begin{tabular}{@{}c@{}}SasRec \\ (CE baseline)\end{tabular}
& \begin{tabular}{@{}c@{}}SasRec \\ (CL4SRec aug.)\end{tabular}
& \begin{tabular}{@{}c@{}}SasRec \\ (SimSiam)\end{tabular}
& \begin{tabular}{@{}c@{}}BT-SR \\ {[}Ours{]}\end{tabular} \\
\midrule

\multirow{7}{*}{\rotatebox{90}{Yelp}}
 & hr@1    
 & 0.0043 
 & 0.0045
 & \underline{0.0047} 
 & \textbf{0.0049} \\

 & ndcg@5  
 & 0.0088 
 & 0.0102 
 & \underline{0.0108} 
 & \textbf{0.0117} \\

 & ndcg@10 
 & 0.0127 
 & 0.0137 
 & \underline{0.0143} 
 & \textbf{0.0150} \\

 & cov@1   
 & \textbf{0.0233} 
 & \textbf{0.0233} 
 & \underline{0.0225} 
 & 0.0218 \\

 & cov@5   
 & 0.0621 
 & \textbf{0.0658} 
 & \underline{0.0642} 
 & 0.0636 \\

 & cov@10  
 & 0.0911 
 & \textbf{0.1009} 
 & 0.0791 
 & \underline{0.0998} \\

\arrayrulecolor{gray!30}\cmidrule(lr){2-6}\arrayrulecolor{black}

\multirow{7}{*}{\rotatebox{90}{Kindle}}
 & hr@1    
 & 0.0533 
 & 0.0539 
 & \underline{0.0546} 
 & \textbf{0.0565} \\

 & ndcg@5  
 & 0.0726 
 & 0.0731 
 & \underline{0.0740} 
 & \textbf{0.0755} \\

 & ndcg@10 
 & 0.0765 
 & 0.0786 
 & \underline{0.0789} 
 & \textbf{0.0796} \\

 & cov@1   
 & 0.0409 
 & \textbf{0.0439} 
 & 0.0425 
 & \underline{0.0433} \\

 & cov@5   
 & 0.1492 
 & \underline{0.1602} 
 & 0.1560 
 & \textbf{0.1656} \\

 & cov@10  
 & 0.2346 
 & \underline{0.2565} 
 & 0.0785 
 & \textbf{0.2665} \\

\bottomrule
\end{tabular}
}
\label{tab:abl}
\end{table}

\subsection{Ablation Study}
\label{sec:ablation}

\paragraph{Impact of Augmentation Strategy.}
We assess the impact of our behaviorally aligned augmentation strategy and provide the results in Table~\ref{tab:abl}. 
Following prior contrastive learning approaches such as CL4SRec~\cite{DBLP:journals/corr/abs-2010-14395}, 
we experiment with perturbation-based augmentations (e.g., random masking, cropping, and reordering). 
However, we observe that these noise-based augmentations lead to performance degradation in our non-contrastive, multi-task setting. 
We further evaluate item-based augmentations on the YELP and Kindle Store datasets and find a performance drop of at least 6\%. 
These findings support our choice of goal-preserving augmentations derived from user intent, as described in Section~\ref{subsec:augs}.
\paragraph{Accuracy–coverage trade-off.}

To analyze the balance between the primary recommendation loss and our Barlow–Twins regularizer, we varied $\alpha$ in the range $[0, 0.1, ...0.5]$, where \(\alpha=0\) effectively removes the BT term in Equation~\ref{eq:total_loss}. We also evaluated sensitivity to the decorrelation weight \(\lambda\) in Equation~\ref{eq:2}. To isolate each effect, we first fixed \(\lambda\) at its optimal value and swept \(\alpha\), and then held \(\alpha\) constant while sweeping \(\lambda\). The plots are available in the GitHub repository (see~\ref{fn:repo}). 

To further dissect this trade‐off, we studied metrics for 3 item-popularity buckets. Figures ~\ref{figure:ml_abol} plot \(\mathrm{hr}@10\) and \(\mathrm{cov}@10\) as functions of \(\alpha\), along with \(\mathrm{hr}@10\) across three popularity buckets relative to the balance point. 
The trade-off parameter \(\alpha\) provides a direct mechanism for navigating between accuracy- and coverage-oriented recommendation regimes, but the effect is dataset-dependent. On YELP, increasing \(\alpha\) beyond the balanced point leads to steadily higher \(\mathrm{cov}@10\), while \(\mathrm{hr}@10\) declines only moderately; at the same time, the bucket-wise analysis shows substantial gains for the least frequent items and mild losses for the most frequent ones. This indicates that stronger regularization shifts recommendation mass toward less frequent items while preserving competitive overall accuracy. We observe the same qualitative behavior on Yelp, Gowalla, and Kindle Store: a low‐\(\alpha\) regime tuned for head‐item precision versus a high‐\(\alpha\) regime that enhances long‐tail coverage (see ~\ref{fn:repo}).
\begin{figure}[!t]
    \centering

    \includegraphics[width=0.5\linewidth]{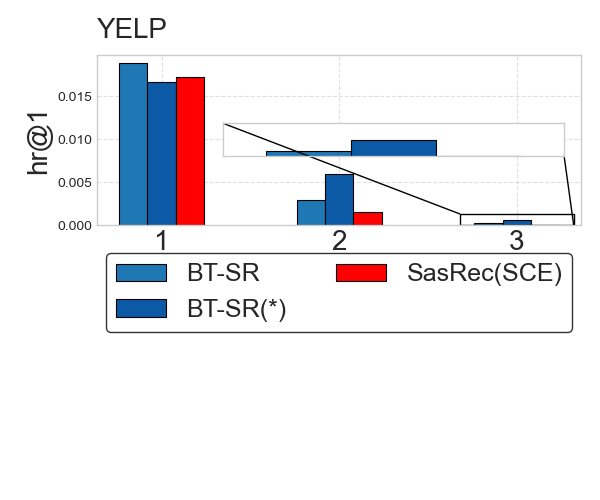}
    \includegraphics[width=0.49\linewidth]{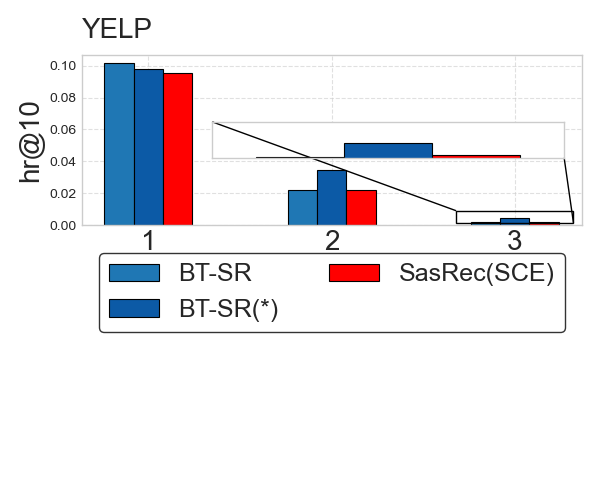}
    \vspace{-4em}
    \caption{Performance comparison on the Yelp dataset under two regimes: BT-SR with $\alpha=0.1$ favors popular items (high HR@1), while BT-SR(*) with $\alpha=0.4$ promotes diverse, less popular items early in the list, boosting personalization and maintaining high overall accuracy (HR@10).}
    \label{figure:pop_buckets_modes}
    \Description[Bar chart compares recommendation metrics on Yelp for BT-SR at $\alpha=0.1$ (head-favoring), BT-SR(*) at $\alpha=0.4$ (tail-favoring) and SASRecc(SCE), across HR@1, and HR@10.]{Figure 4 shows a bar chart for the Yelp dataset comparing two configurations of BT-SR: one with $\alpha=0.1$, which favors popular items and achieves the highest HR@1, and another labeled BT-SR(*) with $\alpha=0.4$, which promotes less popular items earlier in the ranked list, and SASRec(SCE). The chart displays HR@1 and HR@10 side by side for both configurations, showing that the $\alpha=0.4$ setting sacrifices some head-item precision (lower HR@1) but improves personalization and diversity while keeping HR@10 comparably high, demonstrating the practical trade-off enabled by tuning $\alpha$.}

\end{figure}

\paragraph{Dataset-specific effects of $\alpha$.}
On ML-1M, the best $\mathrm{cov}@10$ is achieved at small $\alpha$, while larger values reduce both $\mathrm{hr}@10$ and $\mathrm{cov}@10$. The bucket-wise curves show that this degradation is concentrated in the medium- and low-frequency groups, whereas $\mathrm{hr}@10$ for the most frequent items improves. A likely reason is the interaction between our pair construction strategy and the dataset's timestamp structure. Since BT-SR forms positive pairs from sequences ending with the same target item, frequently occurring items naturally generate more training pairs. In ML-1M, this effect is amplified by timestamp collisions, which make certain items more likely to appear as sequence endpoints \cite{hidasi2023widespread}. As $\alpha$ increases, the regularizer is therefore shaped disproportionately by these frequent terminal items, improving head-item performance while weakening representations for the rest of the catalog.

In Beauty, the effect of $\alpha$ is likely limited by short user histories \cite{klenitskiy2024does}. With small sequential context, sequence representations are driven by only a few items, leaving less room for the regularizer to improve structure. As a result, increasing $\alpha$ brings little benefit and may reduce the model's ability to distribute recommendations broadly, leading to weak coverage gains on this dataset.
\paragraph{Controlling Recommendation Regimes with $\alpha$.} 
Overall, the Figure~\ref{figure:ml_abol} suggests that $\alpha$ acts as a controllable knob that reshapes the allocation of recommendation accuracy across item-frequency groups, with beneficial trade-offs on some datasets and some deterioration on others. We demonstrate the practical aspect of this effect on Figure~\ref{figure:pop_buckets_modes}. We provide two setups of our BT-SR method with two different values of $\alpha$ on the Yelp dataset. Both setups allow outperforming the baseline in terms of integral recommendations accuracy yet they yield different internal structure of recommendations with respect to item popularity. In one regime, corresponding to lower $\alpha$, the recommendations are steered towards more generic user interests, which is indicated by a higher performance in the first bucket in terms of \(\mathrm{hr}@1\) metric. In contrast, the second regime with higher $\alpha$ compensates for lower scores in the first bucket by a better performance in the second and third buckets, thus promoting less popular yet relevant recommendations at the beginning of the recommendation list. It helps increasing the diversity of recommendations without compromising the overall accuracy and boosts personalization.
In practice, we find that moderate values of $\alpha$ (e.g., 0.2–0.4) offer the best balance across datasets.

\section{Conclusion and future work}
We studied feature decorrelation as a mechanism for shaping embedding geometry in sequential recommendation and instantiated this idea as BT-SR by integrating the Barlow Twins redundancy-reduction loss into next-item training. Beyond improving next-item ranking across multiple benchmarks, BT-SR consistently reduces the concentration of shared-subspace alignment across different item-popularity levels, as quantified by our Alignment Concentration diagnostic, supporting the view that decorrelation weakens the low-rank pathways through which particular items may gain an undesired global scoring advantage. 
Importantly, the decorrelation weight provides an explicit control knob that allows practitioners to navigate accuracy--exposure trade-offs in a predictable way, dataset‑dependent way. Overall, our results suggest that non-contrastive redundancy reduction is not only a practical regularizer for sequential recommenders, but also a useful lens for understanding how geometric structure in the embedding space translates into recommendation patterns.

The ability of BT-SR to reshape embedding geometry and reallocate head--tail alignment suggests a promising avenue for controlling popularity-driven concentration, rather than a blind mitigation of popularity bias. The direction for future work is to study this mechanism more directly, identifying which geometric changes are causally responsible for shifts in recommendation exposure across item groups. A deeper analysis of how decorrelation interacts with sequence structure, target-item pairing, and training objectives may help isolate the precise pathways through which representation geometry influences recommendation behavior.


\bibliographystyle{ACM-Reference-Format}
\bibliography{main_bib}

\end{document}